\documentclass[apjl]{emulateapj}

\begin{document}
\title{A Survey of Chemical Separation in Accreting Neutron Stars}
\author{Ryan Mckinven\altaffilmark{1}}
\author{Andrew Cumming\altaffilmark{1}}
\author{Zach Medin\altaffilmark{2}}
\author{Hendrik Schatz\altaffilmark{3}}
\affil{\altaffilmark{1}Department of Physics and McGill Space Institute, McGill University, 3600 rue University, Montreal, QC, H3A 2T8, Canada; ryan.mckinven@mail.mcgill.ca, cumming@physics.mcgill.ca}
\affil{\altaffilmark{2}Los Alamos National Laboratory, Los Alamos, NM 87545, USA; zmedin@lanl.gov}
\affil{\altaffilmark{3}National Superconducting Cyclotron Laboratory and Department of Physics and Astronomy, Michigan State University, East Lansing, MI 48824, USA}

\begin{abstract}
The heavy element ashes of rp-process hydrogen and helium burning in accreting neutron stars are compressed to high density where they freeze, forming the outer crust of the star.  We calculate the chemical separation on freezing for a number of different nuclear mixtures resulting from a range of burning conditions for the rp-process. We confirm the generic result that light nuclei are preferentially retained in the liquid and heavy nuclei in the solid. This is in agreement with the previous study of a 17-component mixture of rp-process ashes by \cite{Horowitz2007}, but extends that result to a much larger range of compositions. We also find an alternate phase separation regime for the lightest ash mixtures which does not demonstrate this generic behaviour. With a few exceptions, we find that chemical separation reduces the expected $Q_{\rm imp}$ in the outer crust compared to the initial rp-process ash, where $Q_{\rm imp}$ measures the mean-square dispersion in atomic number $Z$ of the nuclei in the mixture. We find that the fractional spread of $Z$ plays a role in setting the amount of chemical separation and is strongly correlated to the divergence between the two/three-component approximations and the full component model.
The contrast in $Y_e$ between the initial rp-process ashes and the equilibrium liquid composition is similar to that assumed in earlier two-component models of compositionally driven convection, except for very light compositions which produce nearly negligible convective driving. We discuss the implications of these results for observations of accreting neutron stars. 
\end{abstract}


\section{Introduction}
\label{sec:intro}

In accreting neutron stars, the accreted hydrogen and helium burns a few hours after arriving on the star via the rp-process \citep{Wallace1981,Bildsten1998}. The resulting ashes consist of a complex mixture of heavy elements beyond the iron group \citep{Schatz1998}. These heavy element ashes initially form the liquid ocean of the star but upon further compression freeze to form the outer crust. The composition of the outer crust is an important quantity to understand because it sets the thermal and electrical conductivity, which determines thermal and magnetic evolution (e.g.~\citealt{Brown1998}), and determines the distribution of nuclear heat sources (in particular, in the outer crust, electron captures onto nuclei; \citealt{Haensel2003,Gupta2007}).

\cite{Horowitz2007} showed that chemical separation would occur on freezing of the rp-process ashes. They used molecular dynamics simulations to compute the evolution of a representative 17-component mixture from \cite{Gupta2007}, finding that the lighter nuclei preferentially remain in the liquid phase. \cite{Medin2010} developed a semi-analytic method that was able to closely reproduce these results, building on previous work on one-, two- and three-component plasmas, in particular the work of \cite{Ogata1993} for C--O--Ne mixtures. A comparison of the semi-analytic method with molecular dynamics was also carried out for mixtures of C, O and Ne with application to white dwarf interiors \citep{Hughto2012}.

Chemical separation in neutron star oceans has several observational implications. It can change the concentration and distribution of carbon in the ocean \citep{Horowitz2007}, which is believed to be the fuel for the energetic thermonuclear flashes known as superbursts \citep{Strohmayer2002}. The release of light elements at the base of the ocean will drive convection \citep{Medin2011,Medin2015}, and can change the early lightcurve of cooling neutron stars in transiently-accreting systems that go into quiescence \citep{Medin2014}. Chemical separation can simplify the mixture of elements that are present in the solid outer crust. The degree of scattering of electrons by impurities in the lattice is determined by the impurity parameter $Q_{\rm imp}=\sum_j x_j (Z_j-\langle Z\rangle)^2$, where $Z_j$ is the nuclear charge of species $j$, $x_j$ is the number fraction of species $j$ and the sum is over all species in the mixture. The mean charge is $\langle Z\rangle = \sum_j x_j Z_j$. The impurity parameter can be as large as $\sim 100$ in the ashes of rp-process H/He burning \citep{Schatz1999}. Whether it is significantly reduced by chemical separation at the ocean floor is an important question to resolve.

\cite{Horowitz2007} calculated chemical separation for one particular rp-process ash mixture. However, the rp-process can produce a variety of different compositions depending on burning conditions \citep{Schatz2003}. An open question is to what extent chemical separation occurs for these different mixtures of nuclei. The semi-analytic method of \cite{Medin2010} enables the phase diagram of a mixture to be calculated much more rapidly than a molecular dynamics simulation. We take advantage of this here to calculate chemical separation for a variety of different rp-process ashes. Rather than considering arbitrary mixtures, we use a range of heavy element mixtures resulting from calculations of hydrogen and helium burning involving the rp-process. In \S 2, we calculate the liquid and solid compositions in equilibrium for different initial mixtures. In \S 3, we discuss the implications of our results for observations.

\section{Calculation of the liquid--solid equilibrium for a variety of mixtures}

\subsection{Input mixtures}

We adopt realistic mixtures that result from calculations of rp-process burning under different conditions. An important factor is whether the hydrogen and helium burning is thermally stable or unstable. \cite{Schatz1999} calculated the ashes of stable burning, and showed that the composition for accretion rates $\gtrsim \dot m_{\rm Edd}$ has a range of mean nuclear charge from $\langle Z\rangle\approx 26$--$50$. We focus on the nuclear charge $Z$ rather than mass $A$ because that is the relevant quantity for freezing of a Coulomb liquid. The composition becomes heavier at larger accretion rates, up to a limit at $\approx 50 \dot m_{\rm Edd}$ (where $\dot m_{\rm Edd}$ is the local Eddington accretion rate) beyond which the mixture reaches nuclear statistical equilibrium, driving the mixture back to iron group (this effect can be seen in Figs.~7 and 10 of \citealt{Schatz1999}). \cite{Schatz2003} extended these calculations to lower accretion rates below $\dot m_{\rm Edd}$ and showed that the composition would be much lighter, with $\langle Z\rangle\approx 10$ at $\dot m\approx 0.1\ \dot m_{\rm Edd}$. Unstable burning gives a heavier composition than stable burning because it occurs at a significantly larger temperature, resulting in heavy ashes beyond the iron group with $A\approx 60$--$100$ \citep{Schatz2003,Woosley2004}.

\cite{Stevens2014} recently calculated the rp-process ashes for a large number of steady-state burning models with different accretion rates, base fluxes, and helium fraction in the accreted material. Here we study four models from that paper which have helium-rich accreting material with helium mass fraction $Y$=0.55 and accretion rates $\dot m/\dot m_{\rm Edd}=0.1,0.5,1.0,2.0$ and then a series of models with $Y$=0.2752 and $\dot m/\dot m_{\rm Edd}$ from 0.1 to 40. We also look at the composition used by  \cite{Horowitz2007}, taken from \cite{Gupta2008}, which was the result of unstable H/He burning, and an additional composition resulting from unstable burning. These compositions span a nuclear charge range of $\langle Z\rangle=8$--$35$.

The rp-process ashes are produced well above the ocean floor, at densities $\sim 10^5$--$10^6\ {\rm g\ cm^{-3}}$. We allow for beta decays and electron captures as the ashes move to higher densities by finding the $Z$ for each mass chain $A$ that is in beta equilibrium at an electron Fermi energy $E_F=4\ {\rm MeV}$, which is a typical Fermi energy at the freezing depth (for a one-component plasma, the Fermi energy at the freezing depth is $E_F=1.7\ {\rm MeV}\ T_8 (Z/26)^{-5/3}$, where $T_8=T/10^8\ {\rm K}$). We have checked that our results are not sensitive to the exact choice of $E_F$.

Some of the compositions contain substantial amounts of light elements helium and carbon, which are likely to burn before reaching the base of the ocean where chemical separation occurs. At lower accretion rates the validity of this assumption is much more uncertain and carbon could potentially reach the ocean basin possibly sourcing superbursts \citep{Strohmayer2002}. Recent multi-zone modelling of superbursts find additional crustal heating ($Q_b$) necessary to transition from unstable to stable burning at low accretion rates \citep{Keek2011}. This in conjunction with rp-process ashes with large carbon mass fractions make the assumption of complete and stable burning of carbon even more uncertain at low accretion rates. The interaction of chemical separation with rp-process ashes of these low accretion rates will be discussed further in \S 3. Unless otherwise stated, we assume that light element burning converts the He and C into Mg (Z=12), and so set the He and C number fractions to zero, and add the sum of the original He and C number fractions to the Mg number fraction \footnote{This is an approximation for complete burning of C and He to Mg. In reality, complete burning would imply the abundances of C and He be added to that of Mg. In terms of number fraction this can be written as, $x_{Mg}^{new}=x_{Mg}^{old}+x_{C}/2+x_{He}/6$. The difference between these two reformulations have been studied and are insignificant to the general features of chemical separation considered in this paper.}.

\begin{deluxetable*}{cccccccccccccc}
\tablecaption{Properties of the liquid--solid equilibrium compositions\label{tab:results}} 
\tablehead{\colhead{$\dot{m} / \dot{m}_{\rm Edd}$} & \colhead{$\langle Z\rangle_{i}$} & \colhead{$\langle Z\rangle_{l}$} & \colhead{$\langle Z\rangle_{s}$} & \colhead{$Q_i$} & \colhead{$Q_l$}  & \colhead{$Q_s$} & \colhead{$Y_{e,i}$} & \colhead{$Y_{e,l}$} & \colhead{$Y_{e,s}$} & \colhead{$Y_{e,l} - Y_{e,i}$} & \colhead{$\Gamma_{i}$} & \colhead{$\Gamma_{l}$} & \colhead{$\Gamma_{s}$}}
\startdata
& & & & & & Y=0.2752 & & & & & &\\
\hline
0.1 &	11.4 & 11.6 & 11.2 & 8.3 & 13.5 & 3.1 & 0.4600 & 0.4592 & 0.4609 & -0.0008 & 286.5 & 299.5 & 273.5\\
0.2 & 16.4 & 13.2 & 19.7 & 41.2 & 23.1 & 38.4 & 0.4483 & 0.4581 & 0.4419 & 0.0098 & 509.3 & 348.4 & 670.2 \\
0.3 & 18.8 & 12.8 & 24.8 & 48.1 & 19.1 & 4.6 & 0.4460 & 0.4655 & 0.4367 & 0.0194 & 556.7 & 288.8 & 824.5\\
0.4 & 21.6 & 16.9 & 26.3 &	51.2 & 50.1 & 8.2 & 0.4434 & 0.4582 & 0.4343& 0.0148 & 351.5 & 240.6 & 462.2\\
0.5 &	22.5 & 18.3 & 26.8 & 49.8 & 55.3 & 8.7 & 0.4405 & 0.4535 & 0.4319 & 0.0130 & 318.4 & 232.8 & 404.0\\
0.6 &	23.1 & 19.1 & 27.1 & 49.7 & 58.3 & 8.8 & 0.4388 & 0.4507 & 0.4307 & 0.0119 & 305.3 & 229.4 & 381.2\\
0.7 &	23.4 & 19.4 & 27.5 & 50.6 & 60.4 & 7.9 & 0.4380 & 0.4496 & 0.4302 & 0.0116 & 303.1 & 228.4 & 377.8\\
0.8 &	23.7 & 19.6 & 27.7 & 51.8 & 62.3 & 8.0 & 0.4376 & 0.4490 & 0.4300 & 0.0114 & 301.4 & 227.4 & 375.3\\
0.9 & 23.8 & 19.8 & 27.9 & 52.7 & 64.0 & 8.2 & 0.4372 & 0.4484 & 0.4297 & 0.0111 & 298.5 & 226.2 & 370.9\\
1.0 &	24.0 & 20.0 & 28.0 & 52.8 & 65.1 & 8.5 & 0.4368 & 0.4475 & 0.4295 & 0.0107 & 294.6 & 225.1 & 364.0\\
1.26 & 24.4 & 20.5 & 28.3 & 53.1 & 66.9 & 9.3 & 0.4363 & 0.4462 & 0.4294 & 0.0099 & 289.0 & 224.3 & 353.6\\
1.58 & 24.7 & 21.0 & 28.5 & 52.6 & 67.6 & 9.4 & 0.4363 & 0.4455 & 0.4297 & 0.0092 & 285.3 & 224.3 & 346.2\\
2.0 & 25.1 & 21.6 & 28.6 & 51.1 & 67.2 & 10.8 & 0.4363 & 0.4443 & 0.4304 & 0.0081 & 278.5 & 224.2 & 332.7\\
2.51 & 25.5 & 22.3 & 28.7 & 48.4 & 65.1 & 11.3 & 0.4362 & 0.4433 & 0.4309 & 0.0070 & 272.3 & 224.2 & 320.3\\
3.16 & 26.0 & 23.2 & 28.7 & 44.1 & 61.2 & 12.0 & 0.4357 & 0.4414 & 0.4312 & 0.0057 & 260.0 & 221.3 & 298.6\\
4.0 &	26.7 & 24.5 & 28.8 & 38.1 & 54.0 & 13.0 & 0.4351 & 0.4394 & 0.4316 & 0.0042 & 245.6 & 217.6 & 273.6 \\
5.0 &	27.5 & 25.9 & 29.2 & 34.1 & 48.6 & 14.1 & 0.4334 & 0.4357 & 0.4315 & 0.0022 & 234.1 & 214.2 & 253.9 \\
6.31 & 28.7 & 27.4 & 30.0 & 32.4 & 45.6 & 16.0 & 0.4311 & 0.4320 & 0.4304 & 0.0008 & 253.8 & 247.0 & 260.5\\
8.0 &	30.1 & 29.1 & 31.0 & 32.1 & 42.3 & 20.1 & 0.4300 & 0.4306 & 0.4294 & 0.0006 & 232.2 & 222.0 & 242.4\\
10.0 & 31.3 & 30.5 & 32.1 & 35.5 & 42.0 & 27.7 & 0.4292 & 0.4300 & 0.4285 & 0.0007 & 241.2 & 232.1 & 250.3\\
12.59 & 32.5 & 31.6 & 33.4 & 43.5 & 47.3 & 37.9 & 0.4278 & 0.4288 & 0.4268 & 0.0010 & 252.1 & 240.9 & 263.4\\
15.85 & 33.8 & 32.8 & 34.7 & 52.1 & 53.3 & 49.0 & 0.4263 & 0.4275 & 0.4253 & 0.0011 & 254.9 & 243.4 & 266.4\\
20.0 & 34.9 & 33.9 & 36.0 & 45.9 & 47.5 & 42.0 & 0.4242 & 0.4252 & 0.4233 & 0.0009 & 254.0 & 242.1 & 265.8\\
25.0 & 36.6 & 35.2 & 37.9 & 42.8 & 44.4 & 37.4 & 0.4214 & 0.4223 & 0.4206 & 0.0009 & 251.2 & 236.1 & 266.3\\
30.0 & 25.4 & 22.3 & 28.6 & 48.4 & 65.3 & 11.5 & 0.4366 & 0.4434 & 0.4314 & 0.0068 & 270.1 & 223.0 & 317.1\\
40.0 & 26.0 & 23.4 & 28.6 & 43.3 & 60.7 & 12.3 & 0.4360 & 0.4413 & 0.4318 & 0.0053 & 255.2 & 219.2 & 291.2\\
\hline
& & & & & & Y=0.55 & & & & & & \\
\hline
0.1 & 11.5 & 11.3 & 11.6 & 1.4 & 1.8 & 1.0 & 0.4620 & 0.4623 & 0.4618 & 0.0003 & 199.1 & 195.2 & 203.0\\
0.5 & 13.0 & 13.4 & 12.5 & 18.8 & 25.5 & 11.7 & 0.4569 & 0.4543 & 0.4597 & -0.0025 & 402.7 & 434.3 & 371.0\\
1.0 & 13.8 & 12.2 & 15.4 & 28.0 & 15.9 & 34.9 & 0.4628 & 0.4739 & 0.4543 & 0.0111 & 555.1 & 442.7 & 667.6\\
2.0 & 14.4 & 12.3 & 16.6 & 31.3 & 11.8 & 41.5 & 0.4661 & 0.4833 & 0.4541 & 0.0172 & 590.9 & 434.1 & 747.7\\
\hline
& & & & & & X-ray Bursts & & & & & & \\
\hline
Horowitz \footnote{The nuclear abundances in terms of atomic mass number (A) were not given for this mixture and so corresponding $Y_e$ values could not be calculated.} & 29.3 & 27.7 & 30.9 & 38.9 & 51.4 & 21.0 &  -  & - & - & - & 236.0 & 216.9 & 255.1\\
XRB & 34.5 & 28.6 & 40.4 & 127.5 & 146.5 & 39.1 & 0.4251 & 0.4324 & 0.4201 & 0.0073 & 325.2 & 246.7 & 403.6\\
\enddata
\end{deluxetable*}

\begin{figure}
\begin{center}
\includegraphics[width=1.05\columnwidth]{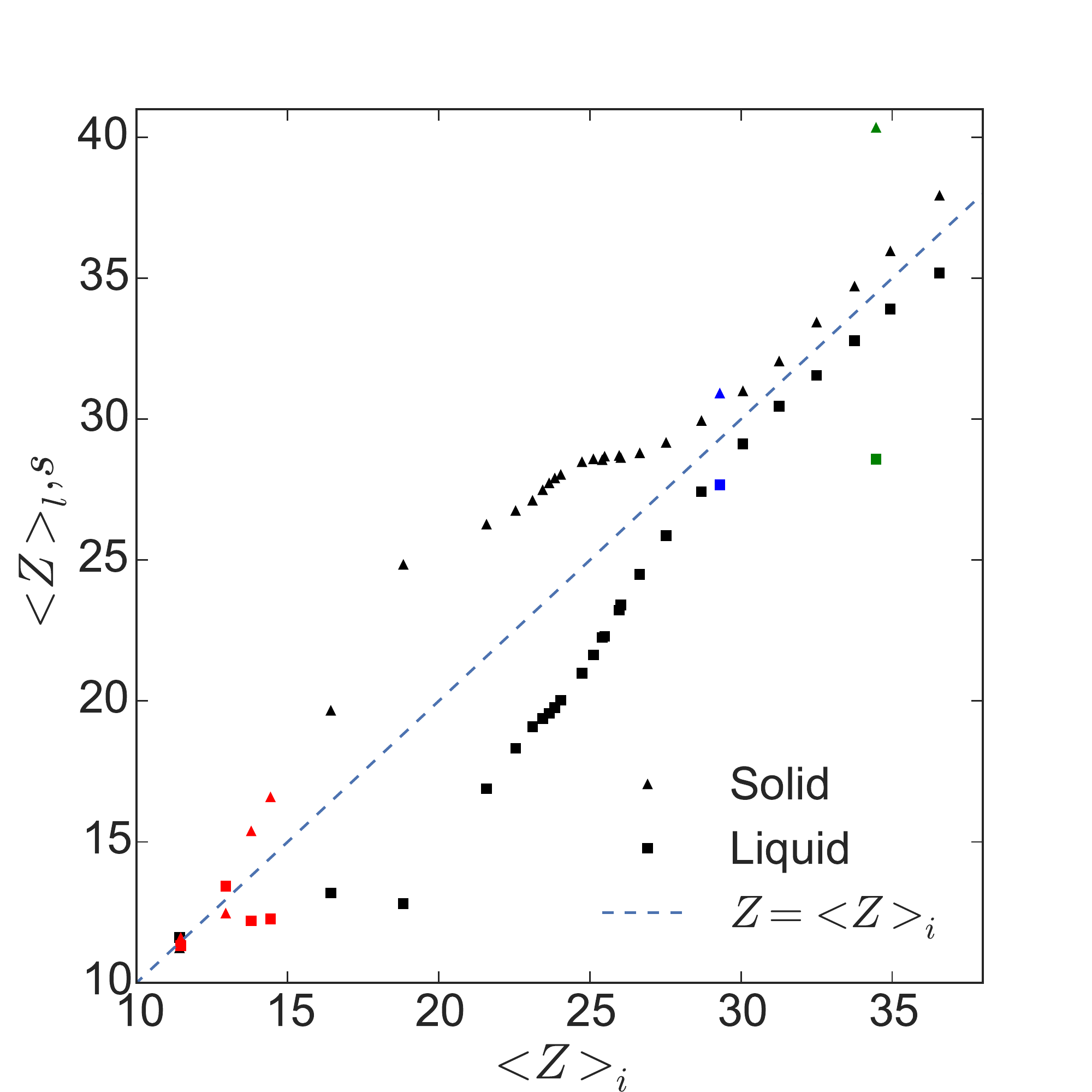}
\end{center}
\caption{The mean charge of the nuclei $\langle Z\rangle$ averaged by number for liquid and solid compositions as a function of $\langle Z\rangle$ of the initial composition. The dashed line shows $\langle Z\rangle_{l,s} = \langle Z\rangle_i$. Black, Red, Blue and Green results correspond to $Y$=$\{$0.2752, 0.55$\}$, \cite{Horowitz2007} and additional X-ray burst data respectively.}
\label{fig:Z}
\end{figure}

\subsection{Calculation of chemical separation}

For each composition, we then use the semi-analytic approach of \cite{Medin2010} to find the composition of the liquid and solid that are in equilibrium with each other. Full details can be found in that paper, but for clarity we give a brief summary of the method here. We start with the linear mixing rule for the free energy of a mixture
\begin{equation}
f^{LM} = \sum_j x_j \left[ f^{OCP}(\Gamma_j)+\ln\left({x_jZ_j\over \langle Z\rangle}\right)\right],
\label{eq:Flm}
\end{equation}
where $f^{OCP}$ is the free energy of a one component plasma, and the logarithmic term is the entropy of mixing. The Coulomb parameter $\Gamma_j$ is given by $\Gamma_j = Z_j^{5/3}\Gamma_e$ with $\Gamma_e=e^2/a_ek_BT$, where $a_e$ is the mean electron spacing $a_e = [3/(4\pi n_e)]^{1/3}$ at the local electron density $n_e$. While the linear mixing rule is adequate to describe the free energy of the liquid phase, the solid phase requires a correction term be included, and so we write the liquid and solid free energies as
\begin{equation}
f_l = f^{LM}_l \hspace{1cm} f_s = f^{LM}_s + \Delta f_s.
\end{equation}
Extending the work of \cite{Ogata1993} on the three-component plasma to an arbitrary number of  $m$ components, \cite{Medin2010} write $\Delta f_s$ as a sum over pairwise interactions,
\begin{equation}
\label{eq:delta_fs}
\Delta f_s = \sum_{j=1}^{m-1}\sum_{k=j+1}^m \Gamma_j x_jx_k \Delta g\left({x_k\over x_j+x_k},{Z_k\over Z_j}\right),
\end{equation}
where the function $\Delta g$ is taken from \cite{Ogata1993}.

Once the free energy is obtained, we then look for an $m-1$ dimensional plane in the $m$ dimensional space of composition that is tangent to the free energy surfaces for the liquid and solid. This $m$-dimensional version of the usual tangent construction then allows one to decompose any composition that lies between the liquid and solid tangent points into coexisting phases of that liquid and solid.

We calculate the liquid and solid compositions that are in equilibrium for a mixture of 50\% liquid and 50\% solid. We calculate the chemical separation using the 17 most abundant species from each rp-process mixture. The choice of 17 species is for convenience, since it matches the number of species in the composition used by \cite{Horowitz2007} and \cite{Medin2010}. However, we have checked that using a smaller or larger number of species in the calculation of chemical separation does not significantly change the results.

\subsection{Results}

The results for all mixtures studied in this paper ($Y=\{$0.55, 0.2752$\}$ and two X-ray burst compositions) are summarized in Table \ref{tab:results} which gives the mean charge $\langle Z\rangle$ and impurity parameter in the initial mixture, liquid and solid for each case. The mean charge is plotted in Figure \ref{fig:Z}. In agreement with the results of \cite{Horowitz2007} for a single mixture, we find the general result that $\langle Z\rangle$ increases for the solid and decreases for the liquid relative to the initial composition \footnote{The notable exceptions to this rule, described further on in the text, are the compositions corresponding to $\dot m$=0.1 $\dot m_{\rm Edd}$, Y=0.2752 and $\dot m$=0.5 $\dot m_{\rm Edd}$, Y=0.55 which produce a `heavier' equilibrium liquid than its solid counterpart.}. This implies that the equilibrium liquid and solid phases will be preferentially enriched in lighter and heavier nuclei respectively relative to the initial mixture.

For the same initial mixture used in the molecular dynamics simulations of \cite{Horowitz2007}, we find good agreement with their results (as did \citealt{Medin2010}). \cite{Horowitz2007} found the average nuclear charge, $\langle Z\rangle$, of the initial and the liquid and solid equilibrium phases to be $\langle Z\rangle_i$=29.30,  $\langle Z\rangle_l$=28.04 and  $\langle Z\rangle_s$=30.48 with equilibrium impurity parameters of $Q_i$=38.9, $Q_l$=52.7, $Q_s$=22.3 and $\Gamma$ values of $\Gamma_{i}$=247, $\Gamma_{l}$=233, $\Gamma_{s}$=261. Our semi-analytic method yields $\langle Z\rangle$ values of $\langle Z\rangle_i$=29.3,  $\langle Z\rangle_l$=27.47 and  $\langle Z\rangle_s$=30.9  with impurity parameters of $Q_i$=38.9, $Q_l$=51.4, $Q_s$=21.0 and $\Gamma$ values of $\Gamma_{i}$=236, $\Gamma_{l}$=217, $\Gamma_{s}$=255 (listed as ``Horowitz'' in Table \ref{tab:results}).


\begin{figure}
\begin{center}
\includegraphics[width=1.05\columnwidth]{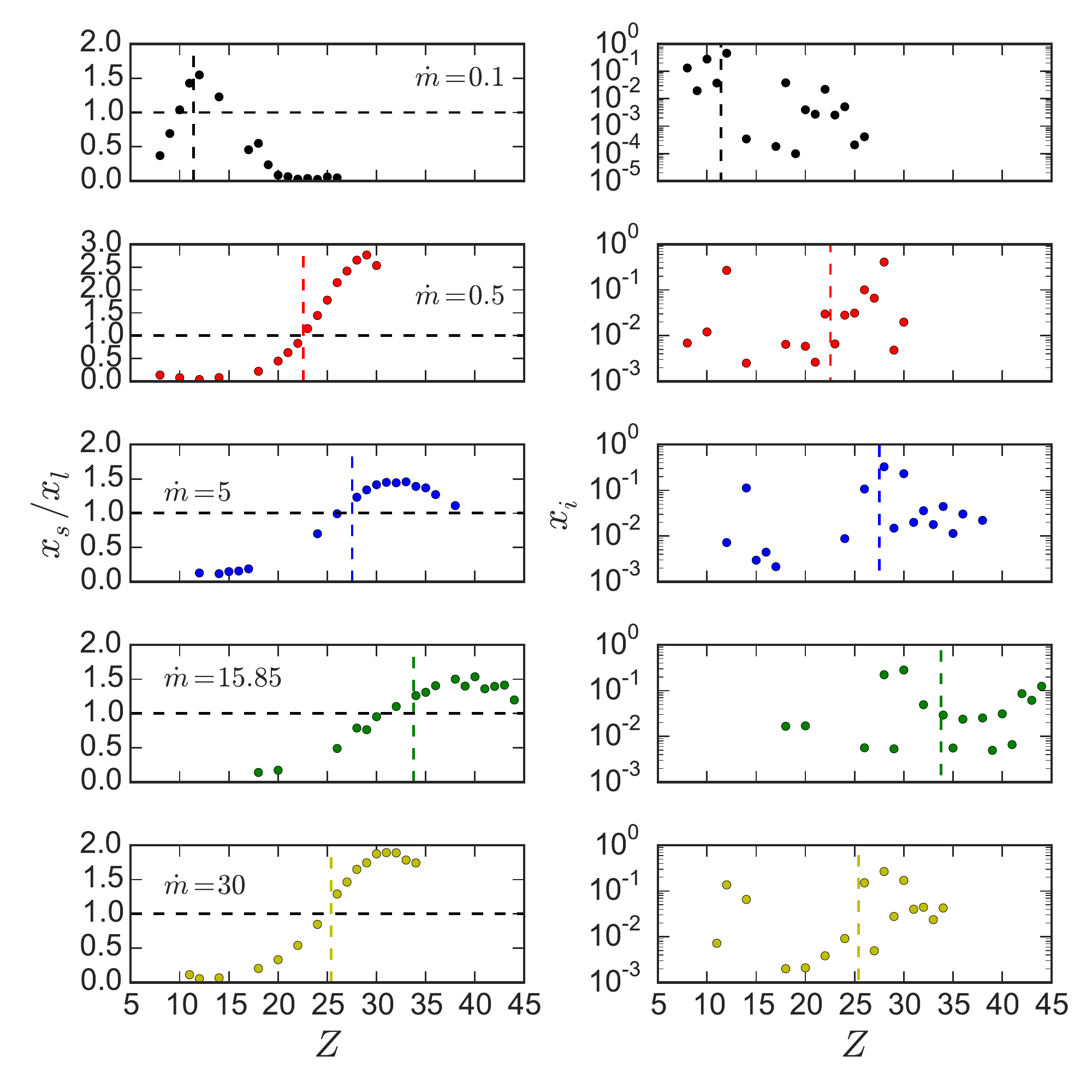}
\end{center}
\caption{The initial number fraction $x_i$ (right panels) and relative number fractions in solid and liquid $x_s/x_l$ (left panels) for five different cases. The models shown are for $Y$=0.2752 and $\dot m=0.1,0.5,5,15.85,$ and $30\ \dot m_{\rm Edd}$. Vertical dashed lines are $\langle Z\rangle_{i}$ of the mixture and horizontal dashed lines mark $x_s/x_l$=1 ``cross-over point". }
\label{fig:abuns}
\end{figure}

The initial number fractions $x_i$ and relative number fractions of solid to liquid $x_s/x_l$ for five different models are shown in Figure \ref{fig:abuns} for helium abundance $Y$=0.2752. Lighter nuclei, relative to the average Z of the composition, are preferentially retained in the liquid ($x_s/x_l<1$), whereas heavier nuclei are found in the solid ($x_s/x_l>1$). This extends the result of \cite{Horowitz2007} to a wider range of compositions. The average atomic number ($\langle Z\rangle_{i}$) of the initial mixture (vertical dashed lines) provides  a quick but imperfect means of determining the ``cross-over point" below which nuclei are preferentially retained in the liquid and above which tend to crystallize in the solid.  This method tends to underestimate the point for lighter compositions ($\langle Z\rangle_{i} <$ 25) and overestimate it for heavier compositions ($\langle Z\rangle_{i} >$ 25). 

\begin{figure}
\begin{center}
\includegraphics[width=1.05\columnwidth]{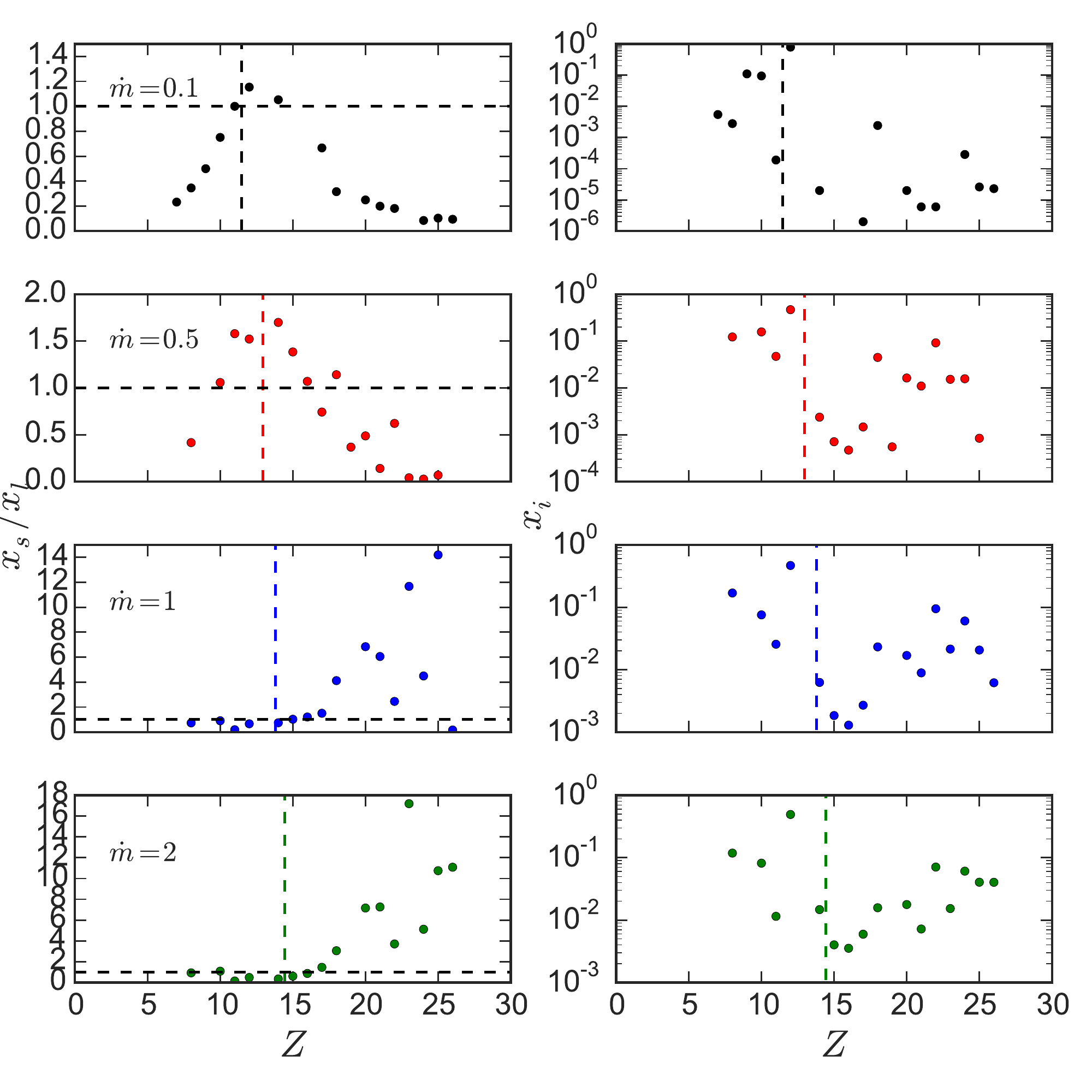}
\end{center}
\caption{The initial number fractions (right panels) and relative number fractions in solid and liquid (left panels) for four different cases. The models shown are for $Y$=0.55 and $\dot m=0.1,0.5,1$ and $2\ \dot m_{\rm Edd}$. Vertical dashed lines are $\langle Z\rangle_{i}$ of the mixture and horizontal dashed lines mark $x_s/x_l$=1 ``cross-over point".}
\label{fig:abunY0.55}
\end{figure}

The top panels of Figure \ref{fig:abuns}, corresponding to an accretion rate $\dot{m}$=0.1 $\dot{m}_{\rm Edd}$, show that for light compositions the general rule of light nuclei being retained in the liquid and heavy nuclei crystallizing in the solid no longer applies. This observation for the light compositions is further verified by equivalent plots for $Y$=0.55 data shown in Figure \ref{fig:abunY0.55}. In this light composition regime ($\langle Z\rangle_{i} < 15$) the correlation of phase with nuclear charge is no longer as strong as for heavier mixtures. Instead, the lightest compositions corresponding to the top panel of Figure \ref{fig:abuns} and the top two panels of Figure \ref{fig:abunY0.55} show that the solid phase is preferentially populated by elements with $Z \sim \langle Z\rangle_{i}$, which includes the most abundant element of the composition (Z=12). These nuclei are not necessarily all heavy relative to $\langle Z\rangle_{i}$, allowing for the possibility of phase separation producing heavier equilibrium liquid than the corresponding solid as is the case for the $\dot m$=0.5 $\dot m_{\rm Edd}$, $Y$=0.55 composition. As the compositions become heavier (i.e. the bottom two panels of Figure \ref{fig:abunY0.55}) the relative number fraction plots ($x_s/x_l$) begin their convergence to the general trend seen in the heavy composition regime.

\begin{figure}
\begin{center}
\includegraphics[width=1.05\columnwidth]{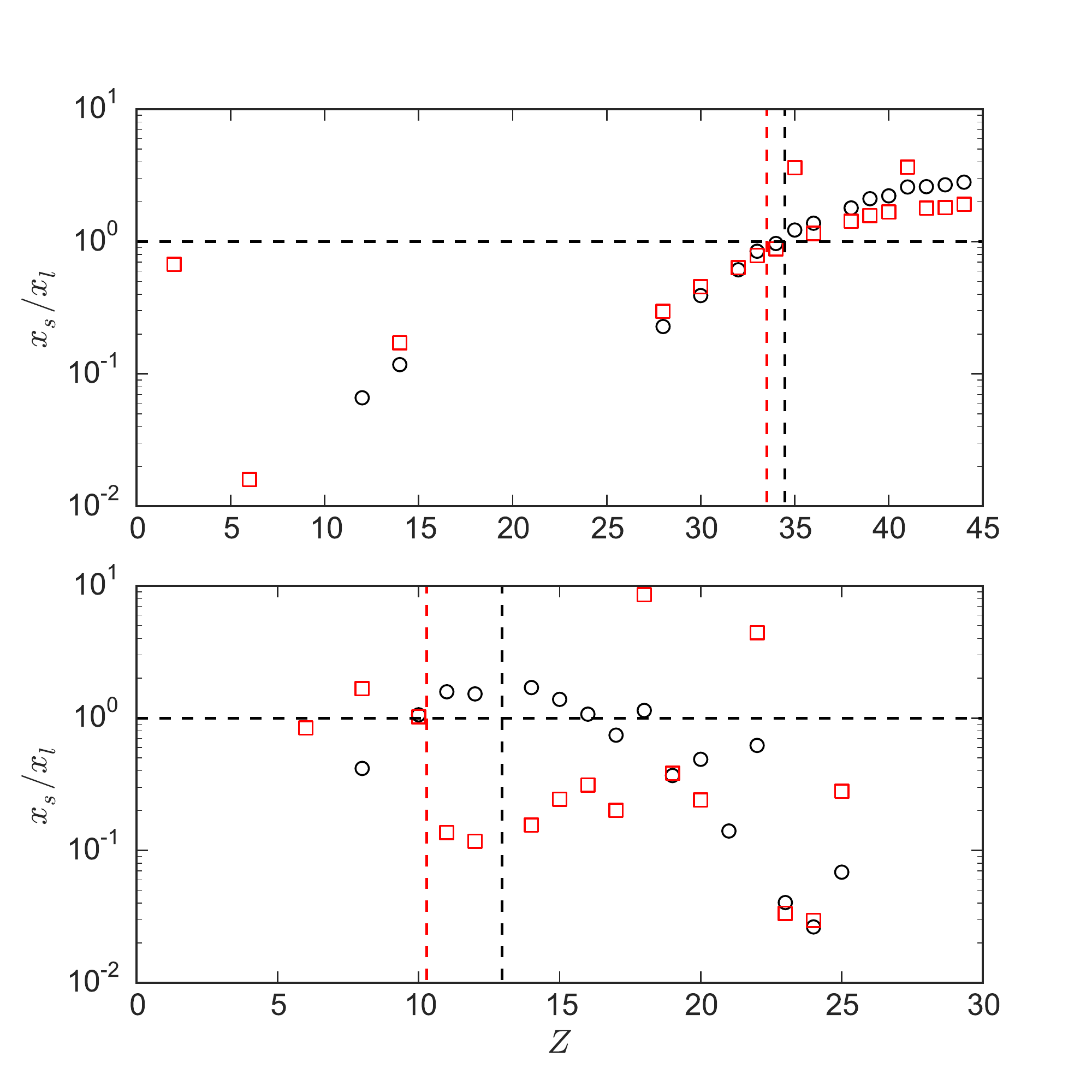}
\end{center}
\caption{The relative number fractions in the equilibrium solid and liquid $x_s/x_l$ for the composition $\dot m=0.5 \dot m_{\rm Edd}$ Y=0.55 (bottom panel) and an X-ray burst ash mixture (top panel). Red data corresponds to compositions assuming no burning of light elements C and He while Black data corresponds to this paper's assumption of complete burning of C and He to Mg. 
\label{fig:carbon}}
\end{figure}

\begin{figure}
\begin{center}
\includegraphics[width=1.05\columnwidth]{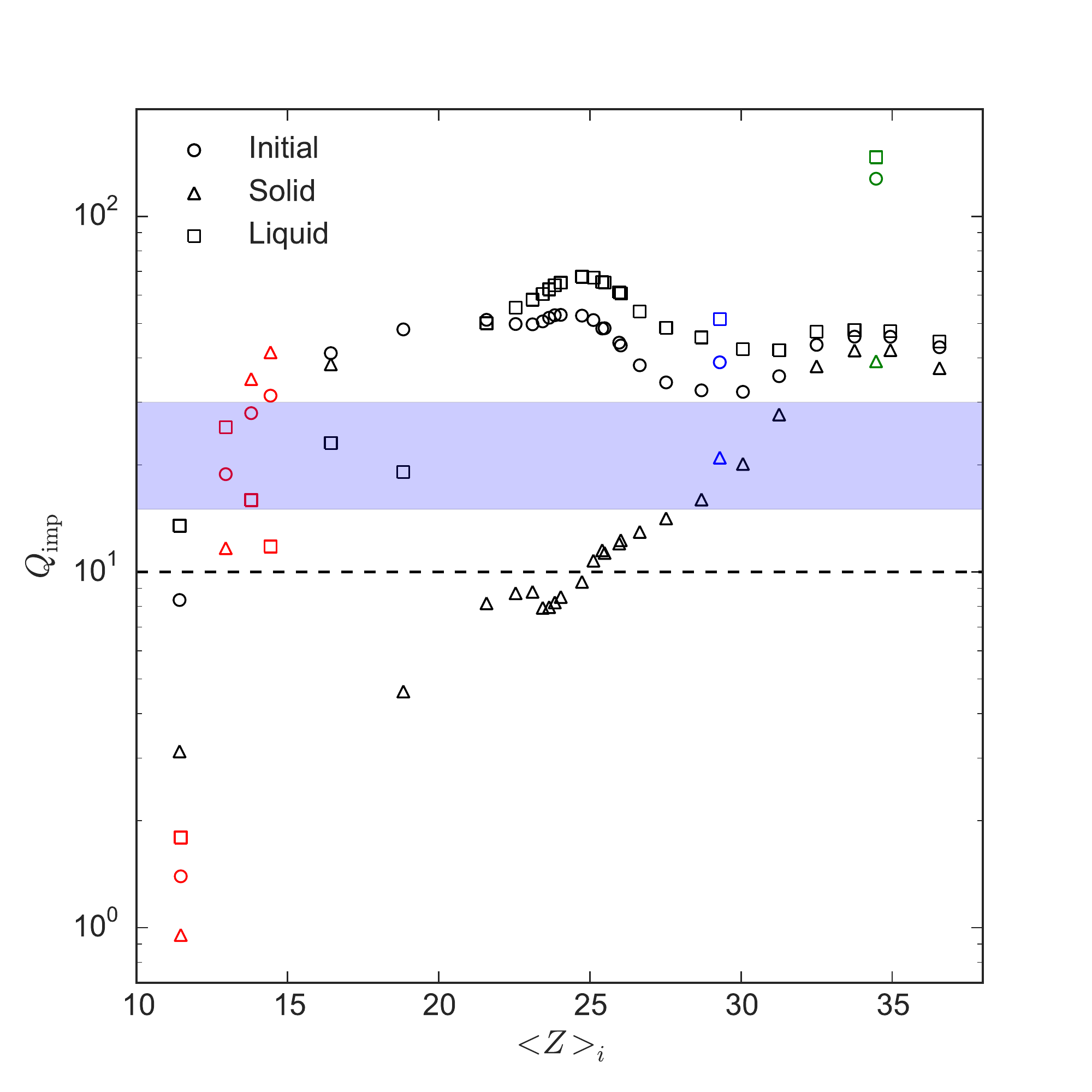}
\end{center}
\caption{Impurity parameter $Q_{\rm imp}$ for the initial mixture, solid, and liquid phases. Chemical separation acts to purify the solid. Black, Red, Blue and Green results correspond to $Y$=$\{$0.2752, 0.55$\}$, \cite{Horowitz2007} and X-ray burst data respectively. The transparent blue band corresponds to the constraints on the outer crust $Q_{\rm imp}$ from \cite{Page2013} from fits to the cooling curve of the transient accreting neutron star XTE~J1701-462. The dotted horizontal line is the upper limit $Q_{\rm imp}<10$ derived by \cite{Brown2009} from fits to the cooling curves of KS~1731-260 and MXB~1659-29. (See \S~\ref{discussion}.)\label{fig:Qimp}}
\end{figure}

If we drop the assumption that He and C burn to Mg before freezing begins, we see a substantial change in the results. The presence of carbon in these light compositions appears to exert a strong influence on the phase separation trend of the $x_s/x_l$ plots. In the light composition regime the rp-process ashes produce substantial amounts of carbon ($x_i^C$=0.42-0.78) which when included with any residual He gives less regular behaviour in the trend of $x_s/x_l$ and tends to magnify $x_s/x_l$ for certain elements. As seen in the bottom panel of Figure \ref{fig:carbon}, for these light mixtures carbon's presence significantly lightens the composition which in turn drastically increases the relative number fractions for specific elements. This effect tends to decrease as the compositions become heavier and the contributions of C and He are no longer as significant. An exception to this was found for the relatively heavy X-ray burst composition (Green data of Figure \ref{fig:Z} and \ref{fig:Qimp}) with its broad range of nuclear species spanning Z=2-44. The top panel of Figure \ref{fig:carbon} shows that including C and He in the composition with initial number fractions ($x_i^C,x_i^{He}$)=(0.044,0.096) causes irregularities to develop in the $x_s/x_l$ trend even though the corresponding $\langle Z\rangle_{i}$ of the composition does not change substantially and is well outside the domain where we expect these irregularities to develop (i.e. outside the light mixture regime). Although these results are interesting it is unlikely that these light compositions with high carbon content reflect the actual composition undergoing phase separation at the base of a neutron star's ocean. These models do indicate, however, that even slight variations in the initial composition can elicit substantial changes to the relative number fractions of certain elements. 

Figure \ref{fig:Qimp} shows the impurity parameter for the initial, solid, and liquid compositions for all mixtures (Red: $Y$=0.55 and Black: $Y$=0.2752). Also plotted are the impurity parameters for the 17-species composition of \cite{Horowitz2007} (Blue) and the additional X-ray burst composition (Green). Across all data sets the impurity parameter of the solid phase is always lower than its initial or liquid counterpart. Comparing Figure \ref{fig:Qimp} with Figure \ref{fig:Z}, we see that the compositions with the largest difference in impurity parameter between liquid and solid, $Q_l-Q_s$, correspond to the compositions with the largest difference in $\langle Z\rangle$ between liquid and solid. In general, a larger enrichment of light elements in the liquid leads to a purer solid phase.

\begin{figure}
\begin{center}
\includegraphics[width=1.05\columnwidth]{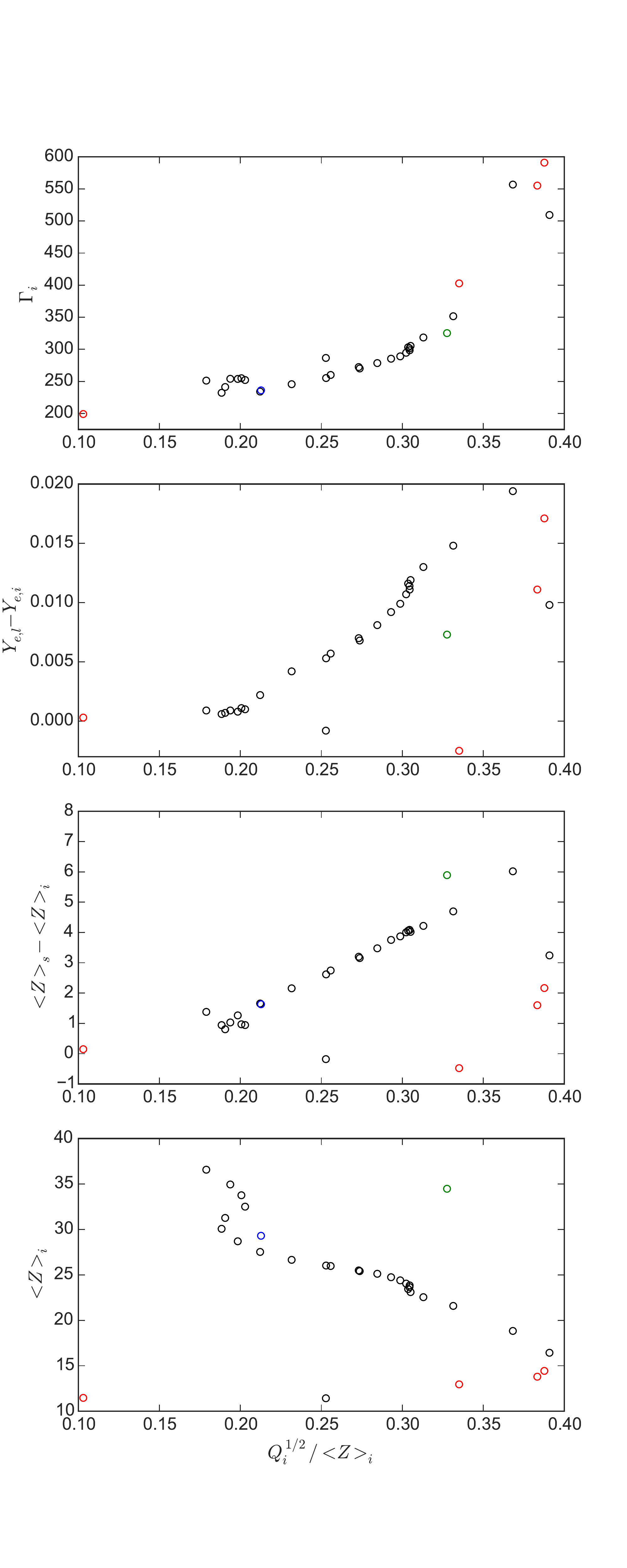}
\end{center}
\caption{The scaling of $\langle Z\rangle_i$, $\langle Z\rangle_s - \langle Z\rangle_i$, $Y_{e,l} - Y_{e,i}$, $\Gamma_i$   with the fractional spread in $Z$ of the initial composition, $\overline{\sigma}=Q_{\rm imp}^{1/2}/\langle Z\rangle_i.$}
\label{fig:std}
\end{figure}

We find that the degree of chemical separation depends on the fractional spread in $Z$ in the mixture, which we measure with the parameter $\overline{\sigma}=Q_{\rm imp}^{1/2}/\langle Z\rangle_{i}$. The bottom panel of Figure \ref{fig:std} shows how this parameter depends on the $\langle Z_i\rangle$ for our mixtures. In general, heavier mixtures tend to have a smaller fractional spread in $Z$ and vice versa, with the exception of the lightest mixtures. The top three panels of Figure \ref{fig:std} show that the Coulomb coupling parameter of the initial mixture at equilibrium ($\Gamma_i$), the contrast in $Y_e$ between liquid and initial mixture ($Y_{e,l}-Y_{e-,i}$), and the heavy element enrichment in the solid ($\langle Z\rangle_{s} - \langle Z\rangle_{i}$) all increase with $\overline{\sigma}$. This means that mixtures with a larger $\overline{\sigma}$ have a lower melting temperature and a greater degree of chemical separation. The lightest mixtures deviate from these trends because they are in the light mixture regime where the solid forms from the most abundant element. In that case, there is no longer a clean separation in $Z$ between elements that go into the liquid and elements that go into the solid, and the degree of chemical separation is much smaller. For example, the models with $\langle Z\rangle_i\leq 13$ in Table 1 have much smaller values of $Y_{e,l}-Y_{e,i}$. These lightest mixtures do still lie on the same trend of $\Gamma_i$ against $\overline{\sigma}$ however, so that $\overline{\sigma}$ is a good predictor of $\Gamma_i$ no matter what the value of $\langle Z\rangle_i$.

\begin{deluxetable}{cccccc}
\tablecaption{Initial number fractions ($x_i$) for the two most abundant species \label{tab:tab1}}
\tablehead{\colhead{$\dot m/\dot m_{\rm Edd}$} & \colhead{$Y$} & \colhead{$Z_1$} & \colhead{$x_1$} & \colhead{$Z_2$} & \colhead{$x_2$} }
\startdata
2 & 0.55 & 12 & 0.49 & 8 & 0.12\\
0.5 & 0.2752 & 28 & 0.41 & 12 & 0.27\\
10 & 0.2752 & 28 & 0.31 & 30 & 0.30
\enddata
\end{deluxetable}

\begin{figure}
\begin{center}
\includegraphics[width=1.05\columnwidth]{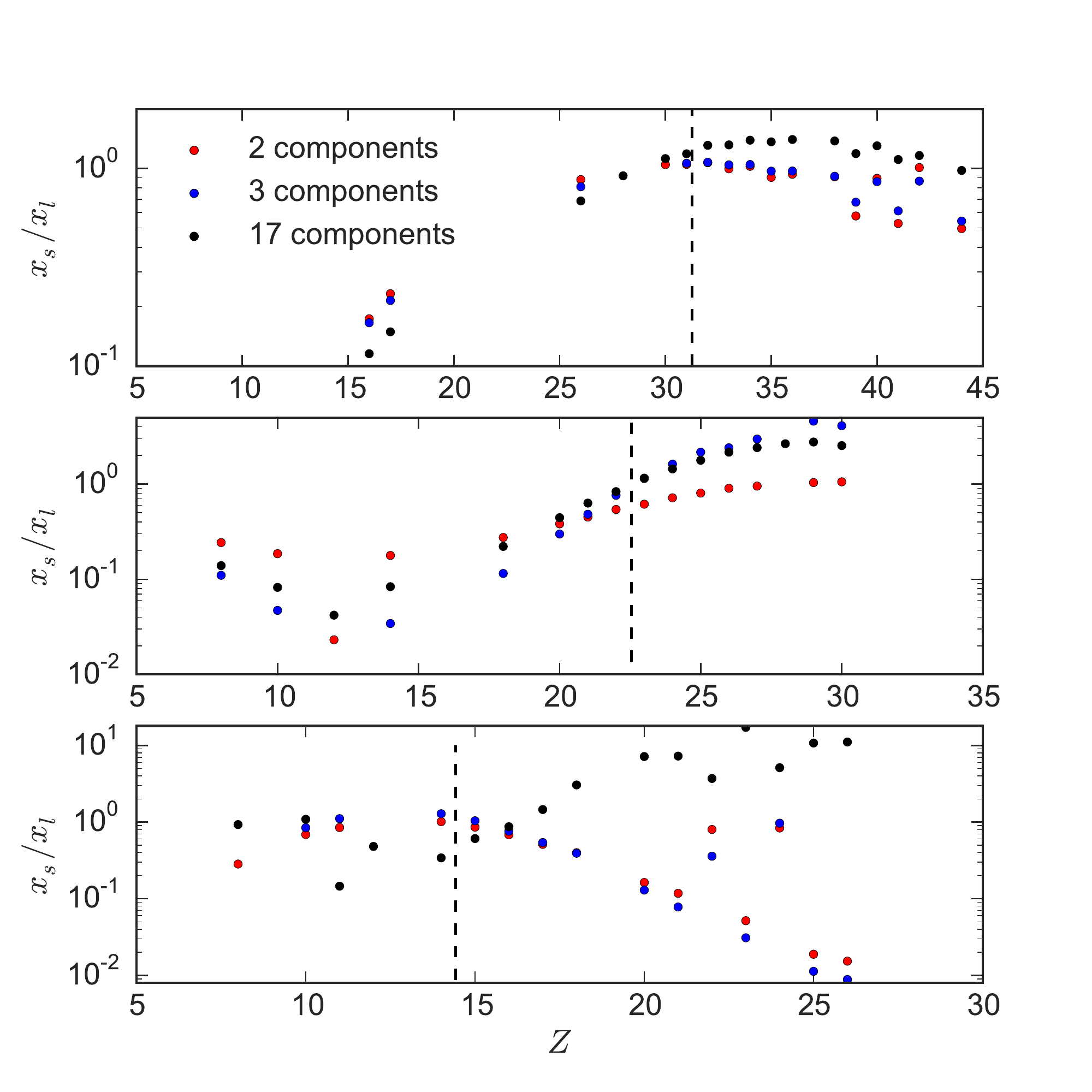}
\end{center}
\caption{A comparison of results from the 17-species calculation to the two/three-component approximation for three steady-state models of $\dot{m}=2$, Y=0.55 (lower panel), 0.5, Y=0.2752 (centre panel) and 10 $\dot m_{\rm Edd}$, Y=0.2752 (upper panel). }
\label{fig:TCP}
\end{figure}

\subsection{The two-component approximation}

\cite{Medin2010} compared their results for the 17-species mixture with a two-component model in which the chemical separation for any given element was calculated by approximating the mixture as consisting of that element plus the most abundant element (which was Se for the 17-species mixture). The pattern of enrichment of heavy elements in the solid and light elements in the liquid was well-reproduced by this simpler model, although with differences in the detailed values of $x_s/x_l$.

Here, we investigate how well the two-component model reproduces our results for a wider range of compositions, and extend it to include three components. Figure \ref{fig:TCP} compares the results from the 17-species calculation to the two- and three-component approximation for three steady-state models summarised in Table \ref{tab:tab1}. For the two-component model, we follow \cite{Medin2010} and calculate the equilibrium solid-to-liquid number fractions for each element assuming the plasma is composed of only two ion species, the element itself and the most abundant element in the mixture. The initial composition of the mixture is chosen such that the ratio of the abundances of the two elements is the same as the 17-component system. In the three-component approximation, we  calculate $x_s/x_l$ of each element assuming the plasma is composed of the two most abundant elements in addition to the element in question. The number fractions are renormalized for each element so that its number fraction relative to the two most abundant elements agrees with the 17-component mixture. 

For most of the compositions the two-component approximation tends to reproduce the general number fraction trend of the 17-component plasma with discrepancies further reduced in the three-component model, as illustrated by the two examples in the upper panels of Figure \ref{fig:TCP}. However, we noticed that compositions with a large $\overline{\sigma}$ show significant differences. In particular, the two and three-component models show the opposite trend of $x_s/x_l$ with $Z$ compared to the 17-component calculation. This can be seen in the lower panel of Figure \ref{fig:TCP}. The difference arises because compositions with large $\overline{\sigma}$ mark the transition between the light and heavy element regimes. In this case, whereas the full composition is close to being in the heavy element regime, the two- and three-component approximations remains firmly in the light elements regime and therefore show the opposite trend.



\section{Discussion} 
\label{discussion}

In this paper, we have applied the semi-analytic method of \cite{Medin2010} to a range of rp-process ash mixtures to survey the extent of chemical separation in accreting neutron stars. We find that for heavy mixtures with $\langle Z\rangle \gtrsim 20$, there is a clean separation between elements with $Z\lesssim \langle Z\rangle$ which are preferentially retained in the liquid, and those with $Z\gtrsim \langle Z\rangle$ which go into the solid (Fig.~\ref{fig:abuns}). The effect is to reduce $Q_{\rm imp}$ from $30$--$60$ in the inital liquid to as low as $10$ in the solid (Fig.~\ref{fig:Qimp}). Lighter mixtures generally show a similar behavior in that the liquid is lighter than the solid ($\langle Z\rangle_l < \langle Z\rangle_s$), but are different in that the preference of any given elements for liquid or solid is harder to predict. For the lightest of these mixtures, the solid phase is dominated by the most abundant element in the mixture which can be either light or heavy relative to  $\langle Z\rangle_i$. The fractional spread in $Z$ as measured by $\overline{\sigma}=Q_{\rm imp}^{1/2}/\langle Z\rangle$ plays a role in determining the amount of chemical separation (Fig.~\ref{fig:std}). The Coulomb coupling parameter $\Gamma_i$, heavy element enrichment as measured by $\langle Z\rangle_{s} - \langle Z\rangle_{i}$, and the contrast in $Y_e$ between liquid and initial mixture all increase with $\overline{\sigma}$. These general behaviours may help in developing simple models of chemical separation to be used in time-dependent calculations of the evolution of accreting neutron stars where it is not feasible or possible to calculate the phase diagram for the complex mixtures of elements that are present. 

\subsection{Possibility of multiple solid phases} 
Our results have implications for the depth of the neutron star ocean, the composition of the outer parts of the neutron star crust, and for driving convection in the neutron star ocean. One caveat is that we have calculated the liquid--solid equilibrium for a 50/50 mixture of liquid and solid, and so have not followed the complete freezing of the mixture. The depth of the ocean floor likely does not correspond to the value of $\Gamma$ for a 50/50 mixture, nor is the composition at the top of the crust the same as the 50/50 equilibrium solid composition. In fact, in steady state, the average composition entering the crust must be the same as the rp-process ashes entering the ocean, and therefore the $Q_{\rm imp}$ in the outer crust would be the same on average as the rp-process ashes.  

It is unlikely, however, that a single phase solid with the same composition as the rp-process ashes can form. For a general multicomponent mixture, multiple solid phases are likely to form instead, with the liquid adopting the composition at the eutectic point. \cite{Horowitz2007} found this for the particular mixture they studied, and a simpler example is shown in Fig.~1 of \citealt{Medin2011} for a two-component mixture of O and Se. The outer crust is therefore likely to consist of multiple solid phases.  Diffusion could result in the different solid phases merging in the outer crust. \cite{Hughto2011} find a diffusion coefficient of $D\sim 10^{-6}\omega_p a^2\sim 10^{-8}\ \rho_9^{-1/6}$ at $\Gamma\sim 180$, just after the solid forms. Since the ocean may take $\sim 100$ years to accrete, the diffusion length is $\sim 3\ {\rm cm}$ close to the top of the crust. This may be enough to merge different solid phases, although the diffusion coefficient decreases sharply with density \citep{Hughto2011}.

We are not able to confidently calculate beyond the 50/50 equilibrium mixture and the possible multiphase solid composition at present, without further comparisons with molecular dynamics simulations. As discussed by \cite{Medin2010}, as the solid fraction increases the resulting compositions depend more sensitively on the particular form of $\Delta f_s$ chosen (eq.~[\ref{eq:delta_fs}]) because $\Delta f_s$ dominates the free energy at large $\Gamma$. It is also not clear if a steady state is reached in the outer crust, as the rp-process ashes are sensitive to variations in accretion rate onto the star. Given these uncertainties, we will assume in exploring the implications of our results that the results we have obtained for 50/50 mixtures give a good estimate of the typical mixtures that make up the outer crust. Further work is needed to constrain the nucleation rate of the equilibrium solid at the ocean-crust boundary. The actual equilibrium mass fraction could therefore differ substantially from the 50/50 mass fraction assumed here.

The equilibrium liquid-solid compositions were also determined for the lightest mixtures (i.e. low accretion rates $\dot{m} \leq 0.5 \dot{m}_{\rm Edd}$) assuming no burning of carbon. We find large variability in the equilibrium $\Gamma_i$ values for these mixtures ($\Gamma_i$=265-626) which roughly equates to an order of magnitude density difference in the approximate location of the ocean-crust boundary. We also find that for the three lightest compositions, the corresponding $\Gamma_i$ values increase with $\langle Z\rangle_i$. This result runs contrary to one-component plasmas which crystallize more readily with increase in nuclear charge. This discrepency is likely due to the increase in the contribution of the entropy of mixing term in equation (\ref{eq:Flm}) as compositions become more heterogenous with increase in accretion rate. Research on the variability of the ocean-crust boundary with composition could offer opportunities to further constrain properties of cooling neutron stars post-outburst or accretion regimes of superbursting sources.
           
\subsection{The impurity parameter in the outer crust}           
The impurity parameter $Q_{\rm imp}$ sets the electron-impurity scattering rate in the crust. Whether this determines the electrical and thermal conductivities depends on how it compares to the electron-phonon scattering rate. At a density $\rho$, impurity scattering will dominate if $Q_{\rm imp}>Q_{\rm crit}$, where 
\begin{equation}
\label{eq:Qcrit}
Q_{\rm crit}=32\ {T_8^2\over \rho_{11}^{5/6}}\left({Z\over 20}\right)\left({Y_e\over 0.4}\right)^{-4/3}
\end{equation}
\citep{Cumming2004}, and $T_8=T/10^8\ {\rm K}$. As density increases, impurity scattering is more likely to dominate, as the increasing Debye temperature reduces the phonon-scattering contribution. 

Constraints on the crust thermal conductivity, and therefore $Q_{\rm imp}$, have come from modelling the observed cooling of neutron star transients in quiescence \citep{Shternin2007,Brown2009,Page2012,Turlione2015}. \cite{Brown2009} assumed a constant $Q_{\rm imp}$ throughout the entire crust and showed that fits to the transient sources MXB~1659-29 and KS~1731-260 required $Q_{\rm imp}\leq 10$, even when uncertainties in the neutron star surface gravity, and therefore crust thickness, were taken into account. This upper limit is indicated by the horizontal dashed line in Figure \ref{fig:Qimp}.

\cite{Brown2009} noted that their constraint on $Q_{\rm imp}$ applies to the inner crust, since electron-phonon scattering sets the conductivity in the outer crust. The value of $Q_{\rm imp}$ in the inner crust likely does not reflect the $Q_{\rm imp}$ of the freshly made solid at the top of the crust. \cite{Steiner2012} modelled the evolution of multicomponent mixtures in the crust as they are compressed and undergo electron captures and neutron emissions and captures near neutron drip. Those models showed a reduction in $Q_{\rm imp}$ near neutron drip, where neutrons are able to redistribute among nuclei (see also \citealt{Gupta2008}). The simplification of the mixture near neutron drip  means that a relatively large $Q_{\rm imp}$ at the top of the crust could be consistent with the constraint $Q_{\rm imp}<10$ from \cite{Brown2009}. Our calculations of the mixture entering the outer crust can be used as input to calculations of the nuclear evolution through neutron drip.

\cite{Page2013} were able to derive constraints on the outer crust $Q_{\rm imp}$. They fit the neutron star transient XTE~J1701-462 using different values of $Q_{\rm imp}$ for the outer and inner crust and found best fitting models with $Q_{\rm imp}\approx 15-30$ at densities $\rho<10^{12}\ {\rm g\ cm^{-3}}$, transitioning to  $Q_{\rm imp}\approx 3-4$ for $\rho>10^{13}\ {\rm g\ cm^{-3}}$. Their constraint for the outer crust is represented by a blue shaded band in Figure \ref{fig:Qimp}. Our results in this paper show that chemical separation on freezing could explain the inferred values of $Q_{\rm imp}$ for the outer crust. Figure \ref{fig:Qimp} shows that the initial values of $Q_{\rm imp}$ are larger than 30 for compositions with $\langle Z\rangle\gtrsim 16$, but that most of the corresponding solid compositions have $Q_{\rm imp}<30$. Further work on the constraints on $Q_{\rm imp}$ in the outer crust from other sources would be interesting.


\subsection{Compositionally-driven convection}
We can use our results to estimate the strength of compositionally driven convection in the neutron star ocean \citep{Medin2011}. The rate of buoyancy production at the ocean--crust boundary depends mostly on the difference in $Y_e$ between the rp-process ash mixture and the equilibrium liquid composition. Electrons dominate the pressure in the ocean (by a factor of $\sim E_F/k_BT\sim 100$), so that the compositional buoyancy is dominated by the gradient of $Y_e$. In that case, we can simplify the expression for the Brunt-V\"ais\"al\"a frequency \citep{BC98,Medin2015} as
\begin{equation}
{N^2H\over g} \approx {\chi_T\over\chi_\rho}\left(\nabla_{ad}-\left.{d\ln T\over d\ln P}\right|_{\star}\right)-{\chi_{Y_e}\over \chi_{\rho}}\left.{d\ln Y_e\over d\ln P}\right|_{\star},
\end{equation}
where we have dropped the ion terms. In this limit, the expression for the convective heat flux in the ocean from \cite{Medin2011} (see their equation~43) becomes
\begin{equation}
\label{eq:Fconv}
F_{\rm conv} = {c_PT\dot m\over \chi_T} \chi_{Y_e} \left({Y_{e,l}-Y_{e,i}\over Y_{e,l}}\right).
\end{equation}
This equation gives the expected heat flux for steady accretion; in a time-dependent situation such as cooling after an accretion outburst, the $\dot m$ factor should be replaced by $\partial y_m/\partial t$, the rate of change of the column depth of the ocean floor due to temperature changes \citep{Medin2015}. Since the pressure is dominated by relativistic degenerate electrons, $\chi_{Y_e}\approx 4/3$, so that the composition enters equation (\ref{eq:Fconv}) through the factor $(Y_e-Y_{e,i})/Y_e$.

The values of $Y_e$ for the initial, liquid, and solid mixtures are given in Table \ref{tab:results}. The largest values of $Y_{e,l}-Y_{e,i}$ are $\approx 0.02$. This is in the range considered by Medin \& Cumming (2011), who looked at two-component mixtures only for simplicity. In that paper, steady state was defined as the point where the composition crystallizing at the bottom of the ocean matched the compostion accreting from the top. This agreement suggests that their results represent a realistic range of outcomes for compositonally-driven convection despite their two-component approximation. It is likely that the values $Y_{e,l}-Y_{e,i}$ tabulated in this paper provide lower bounds on the acual convective heat flux of that mixture. This is because unlike Medin \& Cumming (2011), the results calculated here are for a system where 50 \% of the mixture crystallizes into solid. Generally, this equilibrium solid state is subtantially different (heavier) from the initial mixture accreting at the top of the ocean. Therefore we expect further phase separation and enrichment of the ocean in light elements to arrive closer to the steady state of Medin \& Cumming (2011). Their O--Se mixture, which was chosen to have a similar $\langle Z \rangle$ to the mixture studied by Horowitz et al.~(2007), had 2\% oxygen by mass at the top of the ocean, so that $Y_{e,i}\approx Y_{e,\rm{Se}}=34/79=0.430$. The steady-state composition at the base of the ocean was 37\% oxygen, giving $Y_e-Y_{e,i}=0.026$. Their Fe--Se mixture had $X_{Fe}=0.23$ at the top of the ocean and $X_{Fe}=0.37$ at the base, giving $Y_e-Y_{e,i}=0.005$. Interestingly, some compositions that we study in this paper have $Y_e-Y_{e,i}<0$, so that chemical separation would not lead to convection in those cases. Also we note that the lightest mixtures with $\langle Z\rangle\leq 13$ have smaller values of $Y_{e,l}-Y_{e,i}$ by about a factor of ten compared to the heavier mixtures. Therefore, light element oceans should have much less convective driving.

\subsection{Uncertainties in semi-analytic model} 
Our results rely on an extension of the analytic fits made by \cite{Ogata1993} to Monte Carlo simulations of three-component plasmas to multicomponent plasmas. Even for the three-component case, \cite{Hughto2012} found  systematically lower melting temperatures (higher $\Gamma$) in their MD simulations compared to the semi-analytic model. They noted that this discrepancy seemed to grow with impurity parameter $Q_{\rm imp}$, perhaps suggesting a problem with the form of the deviation from linear mixing $\Delta f_s$ [Equation~(\ref{eq:delta_fs})] assumed in the semi-analytic model. 
Figure \ref{fig:std} shows that $\Gamma_i$ increases with $Q_{\rm imp}$ at fixed $\langle Z\rangle$, which may explain the trend with $Q_{\rm imp}$ seen by \cite{Hughto2011}, since $\Delta f_s$ becomes more and more important at larger values of $\Gamma$. It is important to carry out further comparisons with molecular dynamics simulations \citep{Horowitz2007,Hughto2012} to check and refine these assumptions about the functional form of the free energy
as well as to investigate the formation of multiple solid phases. We hope that our results will give a useful starting point for such comparisons.

\acknowledgements

A.C. is supported by an NSERC Discovery grant, is a member of the Centre de Recherche en Astrophysique du Qu\'ebec (CRAQ), and an Associate of the CIFAR Cosmology and Gravity program. Z.M. recognizes the auspices of the National Nuclear Security Administration of the U.S. Department of Energy at Los Alamos National Laboratory and supported by Contract Nos. DE-AC52-06NA25396 and DE-FG02-87ER40317. H.S. acknowledges support from the National Science Foundation under Grant No. PHY-1430152 (JINA Center for the Evolution of the Elements) and PHY-1102511.

\end{document}